\documentclass[accepted]{uai2026} % for initial submission
\usepackage[british]{babel}

\usepackage{natbib}
\usepackage[utf8]{inputenc} % allow utf-8 input
\usepackage[T1]{fontenc}    % use 8-bit T1 fonts
\usepackage{hyperref}       % hyperlinks
\usepackage{url}            % simple URL typesetting
\usepackage{booktabs}       % professional-quality tables
\usepackage{amsfonts}       % blackboard math symbols
\usepackage{amsmath}
\usepackage{amssymb}
\usepackage{amsthm}
\usepackage{caption}
\usepackage{subcaption}
\newtheorem{proposition}{Proposition}
\newtheorem{corollary}{Corollary}
\newtheorem{remark}{Remark}

\usepackage{nicefrac}       % compact symbols for 1/2, etc.
\usepackage{microtype}      % microtypography
\usepackage{graphicx}
\usepackage{enumitem}
\usepackage{tikz}

\usepackage{algorithm}
\usepackage{algorithmic}
\usepackage[dvipsnames,svgnames,table]{xcolor}
\definecolor{mplblue}{HTML}{1f77b4}
\definecolor{mplorange}{HTML}{ff7f0e}

\def\bo#1{\boldsymbol{#1}}

\def\sBest#1{\textcolor{Gray}{\textbf{#1}}} % Second best

\title{Possibilistic Instrumental Variable Regression \\ with Potentially Invalid Instruments}

\author[1, 2]{Gregor Steiner}
\author[1]{Jeremie Houssineau\thanks{Corresponding author:  \href{mailto:jeremie.houssineau@ntu.edu.sg}{jeremie.houssineau@ntu.edu.sg}}}
\author[2]{Mark F.J. Steel}
\affil[1]{School of Physical \& Mathematical Sciences, Nanyang Technological University, Singapore}
\affil[2]{Department of Statistics, University of Warwick, Coventry, UK}

\begin{document}

\maketitle

\begin{abstract}
Instrumental variable regression is a common approach for causal inference in the presence of unobserved confounding.  However, identifying valid instruments is often difficult in practice. In this paper, we propose a novel method based on possibility theory that performs posterior inference on the treatment effect, conditional on a user-specified set of potential violations of the instrument exogeneity assumption.  Our method can provide valid results even when only a single, potentially invalid, instrument is available. Crucially, and in contrast with existing methods, we prove a finite-sample coverage guarantee for the exactly calibrated (validified) uncertainty intervals when the violation set contains the true value, and we provide practical MC/$\chi^2$ approximations. Simulation experiments and real-data applications indicate strong performance of the proposed approach.
\end{abstract}

%\textbf{Keywords:} Causal inference; Epistemic uncertainty; Invalid instruments; Partial identification; Sensitivity analysis; Unobserved confounding.

\section{Introduction}

Instrumental variables (IVs) offer an approach to estimating treatment effects in the presence of unobserved confounding. An IV is an observed variable that must satisfy three key assumptions:
\begin{enumerate}[label=A\arabic*., itemsep=0pt]
\item The IV is associated with the treatment.
\item The IV is unconfounded, meaning it is not affected by the unobserved confounders.
\item The IV influences the outcome only through its association with the treatment, not directly.
\end{enumerate}
Assumption A1 is known as relevance, while A2 and A3 are typically grouped together and referred to as exogeneity or validity of the instruments. In our linear Gaussian framework, violations of Assumptions A2 and A3 are essentially indistinguishable, but this is not necessarily the case for other model settings. In practice, it is often difficult to find variables that meet all three criteria, and there can be substantial uncertainty about whether candidate instruments are truly valid.

We introduce \underline{v}alid \underline{I}V estimation with \underline{p}ossibilistic \underline{e}ndogeneity \underline{r}obustness (VIPER), a method designed to accommodate slight violations of the exogeneity assumption. If the instruments are not assumed to be valid a priori, there is no one-to-one correspondence between the (always identifiable) reduced-form parameters and the structural parameters of interest. Based on possibility theory \citep{zadeh_fuzzy_1999, dubois_possibility_2015}, our proposed approach can still perform inference on the structural parameters by assigning the ``most-possible'' reduced-form parameters. Because the structural-to-reduced-form map is no longer invertible, we attach to each treatment effect value the plausibility of the reduced-form parameters most compatible with it. We allow the user to specify a set of potential violations and then compute a conditional posterior distribution of the treatment effect given this set. This will generally be uninformative if the set is ``too large'', but can be informative if the violations are small. Unlike many existing methods, our method even supports inference with a single invalid instrument. Based on the possibilistic inferential model construction \citep{martin_inferential_2013, martin_possibilistic_2025}, we prove a frequentist coverage guarantee for the validified uncertainty intervals, which holds if the violation set contains the true value. In practice, we compute plug-in MC or $\chi^2$ approximations to this calibration. %is chosen correctly. 
To the best of our knowledge, no other invalid-IV method provides an analogous finite-sample validity statement for an exactly calibrated plausibility/uncertainty set under a user-specified violation set.

More formally, we study instrumental-variable regression when the instrument may be invalid, and the analyst posits a \emph{violation set} $A$ that contains the direct effect vector $\alpha$. Our main output is a calibrated \emph{possibility function} $\beta \mapsto \pi_w(\beta \mid A)$ that plays the role of a posterior-like uncertainty summary, by conditioning on $\alpha \in A$ for a fixed matrix $w$ of outcomes and treatments, while providing finite-sample frequentist validity. In summary, our contributions are:
\begin{enumerate}[nosep]
\item \textbf{Possibilistic posterior for IV with invalidity sets.} We define a posterior possibility function for the structural parameter $\beta$ conditional on $\alpha \in A$, obtained by profiling over nuisance parameters under a reduced-form Gaussian model.
\item \textbf{Finite-sample calibrated inference via validification.} Using the validification approach of \citet{martin_possibilistic_2025}, we transform the raw profile possibility into a \emph{strongly valid} plausibility function, yielding confidence sets
$C_{\delta}(w, A) = \{\beta : \pi_w(\beta \mid A) \geq \delta\}$ with finite-sample validity (for the exact calibration).
\item \textbf{Computational tractability through convex projection.} For common choices of $A$, evaluating $\pi_w(\beta \mid A)$ reduces to projecting a reduced-form statistic onto $A$, enabling efficient computation and a $\chi^2$ surrogate with an MC reference procedure.
\item \textbf{Sensitivity analysis as a full uncertainty object.} The function $\pi_w(\cdot\mid A)$ supports sensitivity curves over the size/shape of $A$, hypothesis support bounds, and explicit diagnostics of partial identification.
\end{enumerate}

\section{Related work}

\paragraph{Partial Identification.} Allowing for violations of the instrument exogeneity, the treatment effect of interest is only partially identified, meaning multiple values are consistent with the observed data. In such settings, one can typically only estimate bounds on the treatment effect of interest \citep{manski1990nonparametric}. In a seminal paper, \cite{balke_bounds_1997} prove tight bounds on the average treatment effect in a binary experiment with imperfect compliance and unobserved heterogeneity. In contrast, we consider partial identification in a linear structural equation model. Two methods similar to ours are LeakyIV \citep{watson_bounding_2024} and BudgetIV \citep{penn_budgetiv_2025}, where the instruments are allowed to be invalid up to a certain degree. Both LeakyIV and BudgetIV come with confidence intervals that are guaranteed to cover asymptotically, whereas the exact validified VIPER construction yields valid finite-sample coverage if the true degree of misspecification is covered by our violation set.

\paragraph{Plurality rule.} A popular class of methods can perform valid IV estimation if the instruments satisfy the plurality rule \citep{kang_instrumental_2016}. A sufficient condition is that the valid IVs outnumber the invalid ones. Estimation in this setting typically relies on $\ell_1$-penalisation \citep{kang_instrumental_2016, windmeijer_use_2019} or voting/searching strategies \citep{guo_confidence_2018, windmeijer_confidence_2021, guo_causal_2023} to recover the valid instruments. Unlike these approaches, our method is not limited by the plurality rule, but can even perform meaningful inference with no valid instruments, though the inference may be very uninformative.

\paragraph{Sensitivity analysis.} A large strand of the literature focuses on sensitivity analysis where some instruments are allowed to be invalid up to a user-specified degree. \cite{small_sensitivity_2007} focuses specifically on directions where standard overidentifying restrictions tests have low power. \cite{conley_plausibly_2012} propose a partial identification approach that constructs confidence intervals as unions over plausible degrees of misspecification. \cite{andrews2017measuring} investigate the sensitivity of moment-based estimators to misspecified moments. \cite{armstrong_sensitivity_2021} adjust generalised method of moment estimators for misspecification that vanishes asymptotically. \cite{masten2021salvaging} consider set estimates consistent with the smallest relaxation of assumptions such that the model is not falsified. There is a trade-off in the sense that stronger functional form assumptions allow for more robustness to misspecification \citep{deaner2025trade}. \cite{cinelli_omitted_2025} introduce a framework based on robustness values, which describe the minimum strength of association with an instrument needed to overturn conclusions. Our proposed approach can be used in a similar way by gradually widening the violation set.

A wide range of Bayesian (or quasi-Bayesian) approaches exist in the literature, where the sensitivity is specified through a prior distribution \citep[e.g.][]{conley_plausibly_2012}. \cite{chib_bayesian_2018} and \cite{chernozhukov_plausible_2025} use semi-parametric Bayesian methods to analyse a  structural model characterized by a set of moment restrictions, while allowing for the possibility that (some of) these moment conditions do not hold exactly. While \cite{chib_bayesian_2018} focus only on misspecification of overidentifying restrictions, \cite{chernozhukov_plausible_2025} develop a quasi-Bayesian framework to allow for the possibility that all restrictions are invalid, the latter being closer to our framework. \cite{steiner_bayesian_2025} average across different instrument choices based on conditional Bayes factors, providing robustness against falsely using invalid instruments. Most of the approaches mentioned here are probabilistic and widen uncertainty intervals to reflect the additional uncertainty about the instruments' validity. Our contribution is similar, but has the advantage of doing so naturally, as the widened intervals arise directly from epistemic uncertainty about the parameters. Importantly, the exact validified VIPER construction admits a finite-sample coverage guarantee, while other methods are based on asymptotic results, if any.

\paragraph{Alternative uncertainty representations.} Our methodology is grounded in possibility theory \citep{dubois_possibility_2015}, an alternative framework for representing epistemic uncertainty. We perform a modification of Bayesian inference based on possibility measures, which are non-additive outer probability measures \citep{houssineau_elements_2020, hieu_decoupling_2025}. These measures rely on optimisation rather than integration, providing substantial computational benefits in many cases. Finally, the possibilistic inferential model construction \citep{martin_inferential_2013, martin_possibilistic_2025} allows us to transform the raw posterior possibility into a calibrated plausibility function with good sampling properties.

\section{Possibility theory} \label{sec:intro_possibility}

Here, we provide a brief introduction to possibility theory. For more details, we refer to \cite{houssineau_parameter_2020,  houssineau_robust_2022, hieu_decoupling_2025, martin_possibilistic_2025}.

An uncertain variable  $\bo{\theta}$ is a mapping from $\Omega_{\mathrm{u}} \to \Theta$, where $\Omega_{\mathrm{u}}$ is a sample space of deterministic phenomena. We think of $\Omega_{\mathrm{u}}$ as containing a true (but unknown) reference element $\omega_{\mathrm{u}}^*$, thus there is no aleatoric uncertainty connected to $\Omega_{\mathrm{u}}$. We describe the uncertain variable $\bo{\theta}$ by a possibility function $f_{\bo{\theta}} : \Theta \to [0, 1]$ with $\sup_{\theta \in \Theta} f_{\bo{\theta}}(\theta) = 1$. This possibility function gives rise to an outer probability measure
\begin{align*}
    \Bar{\mathbb{P}}_{\bo{\theta}}(A) = \sup_{\theta \in A} f_{\bo{\theta}}(\theta)
\end{align*}
for a subset $A \subseteq \Theta$. Unlike a regular probability measure, $\Bar{\mathbb{P}}_{\bo{\theta}}$ is not additive with respect to disjoint sets. In fact, outer measures can be seen as upper bounds on probability measures. Thus, the possibility $\Bar{\mathbb{P}}_{\bo{\theta}}(A)$ can be interpreted as the maximum subjective probability one would be willing to assign to the set $A$. The possibility function equal to $1$ everywhere on $\Theta$, denoted by $\bo{1}_{\Theta}$, is the most uninformative possibility function in the sense that it assigns full credibility to any (non-empty) set.

Let $\bo{\psi}$ be another uncertain variable on $\Psi$ such that $\bo{\theta}$ and $\bo\psi$ have joint outer measure $$\bar{\mathbb{P}}_{\bo\theta, \bo\psi}(A \times B) = \sup_{\theta \in A, \psi \in B} f_{\bo\theta, \bo\psi}(\theta, \psi), \quad B \subseteq \Psi,$$ where $f_{\bo\theta, \bo\psi}$ is a joint possibility function. Marginalising over $\theta$ is done by setting $A = \Theta$ such that the marginal possibility function of $\bo\psi$ is $f_{\bo\psi}(\psi) = \sup_{\theta \in \Theta} f_{\bo\theta, \bo\psi}(\theta, \psi)$. Conditional outer measures can be defined analogously to probability theory as 
\begin{align*}
    \bar{\mathbb{P}}_{\bo\theta \mid \bo\psi}(A \mid B) = \frac{\bar{\mathbb{P}}_{\bo\theta, \bo\psi}(A \times B)}{\bar{\mathbb{P}}_{\bo\psi}(B)} = \frac{\sup_{\theta \in A, \psi \in B} f_{\bo\theta,\bo\psi}(\theta, \psi)}{\sup_{\psi \in B} f_{\bo\psi}(\psi)}
\end{align*}
as long as $\bar{\mathbb{P}}_{\bo\psi}(B) > 0$, so that the corresponding conditional possibility function is $f_{\bo\theta \mid \bo\psi}(\theta \mid \psi) = f_{\bo\theta,\bo\psi}(\theta, \psi) / f_{\bo\psi}(\psi)$ for all $\psi \in \Psi$ with $f_{\bo\psi}(\psi) > 0$. If $\bo\psi = T(\bo\theta)$ is a transformation of $\bo\theta$, we have that
\begin{align*}
    f_{\bo\psi}(\psi) = \sup \{ f_{\bo\theta}(\theta) : \theta \in \Theta, \psi = T(\theta) \},
\end{align*}
where the appropriate convention is $\sup \emptyset = 0$. There is no need to account for the change in measure by a Jacobian term. This difference to probability theory plays an important role in our proposed methodology.

In this paper, we propose to do Bayesian inference with possibilistic priors. Consider the random variable $Y$ characterised by the statistical model $\{P_\theta : \theta \in \Theta \}$ with corresponding probability density $p(\cdot \mid \theta)$. We incorporate prior information on the parameter $\theta$ in the form of a possibility function $f_{\bo\theta}$. Then, the posterior possibility function is
\begin{align*}
    f_{\bo{\theta} \mid \bo{Y}}(\theta \mid Y) = \frac{p(Y \mid \theta) f_{\bo{\theta}}(\theta)}{\sup_{\theta' \in \Theta} p(Y \mid \theta') f_{\bo{\theta}}(\theta')}.
\end{align*}
The main differences from standard Bayesian inference are that the prior is represented by a possibility function and the denominator is based on maximisation rather than integration. These differences are small enough that much of the intuition from standard Bayesian inference carries through. The possibilistic framework allows vacuous prior information to be modeled by the uninformative possibility function $\bo{1}_{\Theta}$, whereas in standard Bayesian inference an improper prior may yield an improper posterior. It also provides clear computational benefits, since optimisation is generally no harder than integration.

\section{The VIPER methodology}

\subsection{The model}\label{Sec:model}

Let $Y_i$ denote the outcome of interest, $X_i$ a treatment or endogenous variable, and $Z_i$ a $p$-dimensional (row) vector of instrumental variables. We observe $n$ i.i.d. copies of $\{Y_i, X_i, Z_i\}_{i=1}^n$ generated from the structural model
\begin{align} \label{eq:structural_equation}
\begin{aligned}
    Y_i &= \beta X_i + Z_i \alpha + \epsilon_i \\
    X_i &= Z_i \gamma_2 + \eta_i,
\end{aligned}
\end{align}
where the errors are assumed to be jointly Gaussian, $(\epsilon_i, \eta_i)^\intercal \sim N(0, \Sigma)$. Whenever $\Sigma$ is not diagonal, this indicates unobserved confounding (or endogeneity), and ``naive'' inference in the outcome model delivers biased results. In this setting, the instruments $Z_i$ are relevant if $\gamma_2 \neq 0_p$ and exogenous if $\alpha = 0_p$. If $\alpha \neq 0_p$, that indicates correlation between the instrument and the residual $\epsilon_i$ (a violation of A2) or a direct effect of the instruments on the outcome (a violation of A3). Rather than enforcing these assumptions a priori, we incorporate uncertainty about them into the model. In particular, our approach allows for $\alpha$ to be non-zero.

Consider the ``reduced-form'' equation model
\begin{align*}
    (Y_i, X_i) \sim N\left( \begin{bmatrix}
        Z_i \gamma_1 \\
        Z_i \gamma_2
    \end{bmatrix}, \Psi \right),
\end{align*}
where $\gamma_1 = \beta \gamma_2 + \alpha$ and $\Psi$ is the reduced-form covariance given by
$$\Psi = R(\beta) \Sigma R(\beta)^\intercal , \quad R(\beta) =\begin{bmatrix}
    1 & \beta \\ 0 & 1
\end{bmatrix}.$$
Equivalently, we have that the matrix $W = \begin{bmatrix} Y & X \end{bmatrix}$ of stacked outcomes and treatments follows a matrix Normal distribution,  $W \sim MN(Z \Gamma, I_n, \Psi)$, where $Z$ is the matrix with $i$-th row $Z_i$, and the coefficient matrix is $\Gamma = \begin{bmatrix} \gamma_1 & \gamma_2 \end{bmatrix}$.

\begin{remark}
    For simplicity of exposition, we do not explicitly account for exogenous covariates, but these can be easily considered: for a matrix of exogenous covariates $U$, one can project out their effects by premultiplying $W$ and $Z$ by $M_U = I_n - U(U^\intercal U)^{-1}U^\intercal$. This corresponds to marginalising out their effect possibilistically. To see this, consider the extended reduced-form model $W \sim MN(Z \Gamma + U \Delta, I_n, \Psi)$, where $\Delta$ is the exogenous covariates' coefficient matrix. Setting $\Delta$ to $\Delta^*(\Gamma) = (U^\intercal U)^{-1} U^\intercal (W - Z \Gamma)$ maximises the reduced-form likelihood. Thus, under vacuous prior information, plugging in $\Delta^*(\Gamma)$ is the appropriate marginalisation, which yields the model $M_U W \sim MN(M_U Z \Gamma, I_n, \Psi)$.
\end{remark}

The structural parameters are identifiable if we can find a unique solution $(\alpha, \beta, \gamma_2, \Sigma)$ given $(\gamma_1, \gamma_2, \Psi)$, or equivalently, if there exists a bijective mapping between the reduced-form and the structural parameters. The matrix $R(\beta)$ is invertible for any $\beta \in \mathbb{R}$ and $\gamma_2$ maps to itself. Thus, the structural parameters are identifiable if and only if
\begin{align} \label{eq:identification}
    \gamma_1 = \beta \gamma_2 + \alpha \iff \Gamma \begin{bmatrix} 1 \\ -\beta    \end{bmatrix} = \alpha
\end{align}
has a unique solution $(\alpha, \beta)$ given $\Gamma$. Without any assumptions on $\alpha$, this is not the case. Typically, one assumes $\alpha = 0_p$, or that at least the majority of its components are zero \citep{kang_instrumental_2016}.

\subsection{Possibilistic inference}

We propose to perform possibilistic posterior inference %\footnote{See Section \ref{sec:intro_possibility} for a brief introduction to possibility theory.} 
in the reduced-form model and propagate the uncertainty to the structural parameters.  Let $f$ be a prior possibility function on $(\Gamma, \Psi)$, then the posterior possibility $f_{\mathrm{RF}}$ for the reduced form parameters is given by
\begin{multline*}
    f_{\mathrm{RF}}(\Gamma, \Psi \mid W) = \\ \frac{p(W \mid \Gamma, \Psi) f(\Gamma, \Psi)}{\sup_{\Gamma' \in \mathbb{R}^{p\times 2}, \Psi' \in \mathbb{S}^2_{+}} p(W \mid \Gamma', \Psi') f(\Gamma', \Psi')}
\end{multline*}
where $\mathbb{S}^2_{+}$ is the cone of positive-semidefinite and symmetric $2 \times 2$ matrices. Inference is conditional on the instruments $Z$, but we omit this dependence for simplicity. Under the uninformative prior, i.e., $f(\Gamma, \Psi)=1$ for all $\Gamma$ and $\Psi$, the supremum in the denominator is attained by the standard maximum-likelihood estimators.

Then we can define a possibility function $f_{\mathrm{S}}$ for the structural parameters as
\begin{align*}
    & f_{\mathrm{S}}(\alpha, \beta, \Sigma \mid W) = \\ & \sup \left\{ f_{\mathrm{RF}}(\Gamma, \Psi \mid W): \Gamma \begin{bmatrix} 1 \\ -\beta    \end{bmatrix} = \alpha, \Psi = R(\beta) \Sigma R(\beta)^\intercal \right\}.
\end{align*}
This operation does not yield a valid probabilistic posterior on the structural parameters, since the map from reduced-form to structural parameters is not one-to-one in general and the usual change-of-variables formula does not apply. Under uninformative prior possibility functions on the reduced-form parameters, this optimisation problem can be solved in closed form (see Appendix \ref{sec:structural_function}). We have that \( f_{\mathrm{S}}(\alpha, \beta, \Sigma \mid W) = f_{\mathrm{RF}}(\Gamma^*(\alpha, \beta, \Sigma), R(\beta) \Sigma R(\beta)^\intercal \mid W) \), where the optimal reduced-form coefficient matrix is  
\[
    \Gamma^*(\alpha, \beta, \Sigma)
    = \Hat{\Gamma} + \frac{1}{\sigma_{11}}
    \left(\alpha - \Hat{\Gamma} 
    \begin{bmatrix} 1 \\ -\beta \end{bmatrix}\right)
    \begin{bmatrix} 1 & 0 \end{bmatrix}
    \Sigma R(\beta)^\intercal,
\]
with \( \Hat{\Gamma} = (Z^\intercal Z)^{-1} Z^\intercal W \) denoting the least-squares estimate of the reduced-form coefficient matrix, and \( \sigma_{11} \) representing the marginal variance of the outcome in the structural model.

\begin{remark}
    We focus on the case with vacuous prior information, i.e., $f(\Gamma, \Psi) = 1$. More informative priors can be incorporated when additional regularisation is needed, such as in scenarios involving many weak instruments. For instance, a possibilistic Matrix Gaussian prior on $\Gamma$ with column covariance $\Psi$ also leads to a closed-form solution for $\Gamma^*(\alpha, \beta, \Sigma)$. However, the induced prior on $\beta$ is no longer uninformative.
\end{remark}

Our main object of interest is the posterior possibility of $\beta$ given that $\alpha$ lies in a violation set $A$. The following proposition characterises this conditional posterior possibility function.

\begin{proposition} \label{prop:conditional_posterior}
Let \(A \subseteq \mathbb{R}^p\) be the considered violation set. Denote by \((\hat{\gamma}_1, \hat{\gamma}_2)\) and \(\hat{\Psi}\) the maximum-likelihood estimates of the reduced-form coefficients and covariance matrix, respectively, and define $t(\beta) := \hat{\gamma}_1 - \beta \hat{\gamma}_2$. Then, the posterior possibility function of $\beta$ conditional on $\alpha \in A$ is
\begin{align*}
f(\beta \mid \alpha \in A, W)
    = \frac{f_{\mathrm{S}}(\hat{\alpha}(\beta), \beta, \hat{\Sigma}(\beta) \mid W)}{\sup_{\beta' \in \mathbb{R}} f_{\mathrm{S}}(\hat{\alpha}(\beta'), \beta', \hat{\Sigma}(\beta') \mid W)}
\end{align*}
where
\begin{align}
\hat{\Sigma}(\beta) & = R(\beta)^{-1}\, \hat{\Psi}\, [R(\beta)^{\intercal}]^{-1},\\
\hat{\alpha}(\beta) & =\mathrm{Proj}_A^{Z^{\intercal}Z}(t(\beta))
\end{align}
and \(\mathrm{Proj}_A^{Z^{\intercal}Z}\) denotes the projection onto \(A\) with respect to the metric induced by \(Z^{\intercal}Z\). 
\end{proposition}

\begin{proof}
See Appendix~\ref{sec:conditional_beta}.
\end{proof}

The extreme case of $A = \mathbb{R}^p$, where $\alpha$ is completely unconstrained, results in the uninformative marginal possibility function of $\beta$, that is for all $\beta \in \mathbb{R}$ we have that $f(\beta \mid W) = f(\beta \mid \alpha \in \mathbb{R}^p, W) = 1.$ This is not surprising, as $\beta$ is not identified without extra information on $\alpha$, thus we get back the prior. More generally, the intersection of the affine subspace \(t(\beta)\) and the violation set \(A\) defines the partial identification region, in which all values of \(\beta\) are equally plausible. Figure~\ref{fig:geometric_illustration} illustrates this in two dimensions. To obtain informative results, \(A\) must be sufficiently restrictive so that, for most values of \(\beta\), the implied \(\alpha\) lies outside this region.

If $A$ is a rectangle, computing the projection is a standard quadratic programming problem. Alternatively, we can also bound a norm of $\alpha$, that is, specify the constraint set as  $A_\tau = \left\{ \alpha : \lVert \alpha \rVert \leq \tau \right\}$, where the threshold $\tau$ is the maximum invalidity budget across all instruments \citep[similar to][]{penn_budgetiv_2025}. More details are provided in Appendix~\ref{sec:conditional_beta}. In either case, it is important that the choice of $A$ corresponds to the scale of the instrument data. To simplify the interpretation, it may be useful to standardise the instruments. On a common scale, one would typically expect a direct effect of the instruments on the outcome to be smaller than their effect on the treatment, which can serve as a starting point for choosing $A$.

\begin{figure}
    \centering  
    \begin{tikzpicture}[scale=3]
    
      % Axes with labels centered around origin
      \draw[->] (-1,0) -- (1,0) node[right] {$\alpha_1$};
      \draw[->] (0,-0.6) -- (0,1) node[above] {$\alpha_2$};
    
      % Coordinates of line endpoints
      \coordinate (A) at (-0.8,0.85);
      \coordinate (B) at (0.8,0.05);
    
      % Line segments
      \draw[mplblue, thick] (A) -- (B);
    
      % Label near start of line
      \node at (A) [above, mplblue] {$t(\beta) = \hat{\gamma}_1 - \beta \hat{\gamma}_2$};

      % Transparent shaded rectangle covering origin
      \fill[fill=mplorange, fill opacity=0.3] (-0.4,-0.4) rectangle (0.4,0.4);
      \node[mplorange] at (-0.4,-0.4) [above right] {$A$};
    
       % From A to closest rectangle boundary point (-0.4,0.8) vertically above B
       \filldraw[mplblue] (-0.35, 0.625) circle (0.01) node[above right] {$\alpha$};
       \filldraw[mplorange] (-0.35, 0.4) circle (0.01) node[below right] {$\alpha^*$};
       \draw[->, thick, mplorange] (-0.35, 0.625) to[out=200, in=150] (-0.35,0.4);
    
    \end{tikzpicture}
    \caption{\textbf{A geometric illustration of our method for $p=2$:} The causal effect $\beta$ is partially identified where the affine subspace $t(\beta) = \Hat{\gamma}_1 - \beta \Hat{\gamma}_2$ intersects the tolerated region $A$. At these values of $\beta$, the corresponding $\alpha$ is precisely $t(\beta)$, and therefore, the conditional possibility is $1$. For all other values of $\beta$, the optimal $\alpha$ is the projection onto $A$ with respect to the metric induced by $Z^\intercal Z$.}
    \label{fig:geometric_illustration}
\end{figure}

Our procedure can be viewed from two different perspectives. The first treats the choice of \(A\) as partial prior information based on domain knowledge, specified by the analyst before observing the data. Incorporating this prior information through post-hoc conditioning, rather than directly specifying a prior possibility function, is convenient for two reasons: (i) it allows closed-form solutions for many of the expressions, and (ii) specifying a prior on the reduced-form parameters that induces the desired prior on the structural parameters is challenging. The second perspective emphasises sensitivity analysis for a particular effect, where the analyst gradually widens \(A\) to assess how much instrument invalidity would be required for the effect to disappear.

\subsection{Validification}

The sampling properties of our posterior possibility can be improved by using the validification procedure proposed by \cite{martin_inferential_2013}. Specifically, we transform the posterior possibility defined in Proposition \ref{prop:conditional_posterior} to the validified posterior possibility function
\begin{align} 
    & \pi_w(\beta \mid A) = \nonumber \\
    \label{eq:validified_posterior}
    & \quad \sup_{\theta \in \Theta_A} P_{\theta} \left( f(\beta \mid \alpha \in A, W) \leq f(\beta \mid \alpha \in A, w) \right),
\end{align}
where \(w\) denotes the observed value of \(W\), \(P_\theta\) represents the probability measure of \(W\) as a functional of the structural parameters $\theta = (\alpha, \beta, \Sigma, \gamma_2) \in \Theta = \mathbb{R}^p \times \mathbb{R} \times \mathbb{S}^2_{+} \times \mathbb{R}^p$, and $\Theta_A = \{(\alpha, \beta, \Sigma, \gamma_2) \in \Theta: \alpha \in A\}$ is the restriction to the violation set. The supremum is necessary as $\beta$ and $A$ alone do not fully characterise the distribution of $W$. The following proposition shows that the validified posterior possibility controls the type-I error, that is, its probability of assigning ``too little'' possibility to the true value of $\beta$ is bounded at the nominal level, as long as the violation set $A$ contains the true value of $\alpha$.

\begin{proposition} \label{prop:validity}
    Assume the violation set $A$ contains the true value of $\alpha$. Then, the validified conditional posterior possibility $\pi_w(\cdot \mid A)$ as defined in (\ref{eq:validified_posterior}) is strongly valid in the sense that for any $\delta \in [0, 1]$
    \begin{align*}
        \sup_{\theta \in \Theta_A} P_\theta \left( \pi_W(\beta \mid A) \leq \delta \right) \leq \delta,
    \end{align*}
    where $\beta = \beta(\theta)$ is the true $\beta$ for a given $\theta \in \Theta_A$.
\end{proposition}
\begin{proof}
    See Appendix \ref{sec:proof_validity}.
\end{proof}

An immediate corollary is that the upper level sets are valid confidence sets.

\begin{corollary}
    Assume the violation set $A$ contains the true value of $\alpha$. Then, for any $\delta \in [0, 1]$, the upper $\delta$ level set of $\pi_w(\cdot \mid A)$,
    $$
    C_\delta(w, A) = \{\beta \in \mathbb{R}: \pi_w(\beta \mid A) \geq \delta \},
    $$
    is a valid $100(1-\delta)\%$ confidence set, i.e., $\inf_{\theta \in \Theta_A} P_\theta \left( \beta \in C_\delta(W, A) \right) \geq 1 - \delta$.
\end{corollary}

Our inference is monotone with respect to $A$ in the sense that for $A_1 \subseteq A_2$ we have $\pi_w(\beta \mid A_1) \leq \pi_w(\beta \mid A_2)$ pointwise in $\beta$. Therefore, $A$ must be chosen large enough to contain the true value of $\alpha$ to achieve type-I error control. At the same time we want to maximise efficiency (minimise the type-II error), which is inversely proportional to the volume of $A$. Thus, one faces a trade-off between choosing $A$ large enough for it to likely contain the true $\alpha$, but not choosing it too large so that the inference becomes overly conservative. In practice, this choice should be based on domain knowledge. When sensitivity analysis is the goal, one may be interested in finding the largest set $A$ such that a particular effect holds (e.g., such that $0 \notin C_\delta(w, A)$). For instance, starting from $A_0 = \{0\}$ and widening towards larger admissible violations, one can report the largest degree of instrument invalidity at which a conclusion of interest still holds. This plays a similar role to the robustness values of \citet{cinelli_omitted_2025}, reporting in interpretable units how much exogeneity violation a finding can tolerate. However, it is not obvious how to trade-off violations in different directions with multiple instruments. We leave a detailed treatment of this for future work.

The validified posterior possibility \(\pi_w(\cdot \mid A)\) is not directly available, but a natural Monte Carlo approximation is
\begin{align*}
    & \pi_w(\beta \mid A) \approx \\
    & \frac{1}{M} \sum_{j=1}^M 1\{ f(\beta \mid \alpha \in A, W = W_j) \leq f(\beta \mid \alpha \in A, W = w) \},
\end{align*}
where $W_j$ are independent samples from the plug-in measure $P_{\hat{\theta}(\beta, A)}$ based on the maximisers of Proposition~\ref{prop:conditional_posterior} for given $\beta$ and $A$. This plug-in strategy yields an approximately calibrated approximation to the (strongly valid) validified possibility in \eqref{eq:validified_posterior} \citep[][Section 5]{martin_possibilistic_2025}, but is computationally expensive. A cheaper approximation is the Wilks' style $\chi^2$ approximation
$$\pi_w(\beta \mid A) \approx 1 - F(-2 \log f(\beta \mid \alpha \in A, W = w)),$$
where $F$ is the cdf of a $\chi^2$ random variable with $1$ degree of freedom. This approximation is based on asymptotic Gaussianity and is less appropriate whenever the parameter is not point-identified, i.e., when $A$ is not a singleton. For a comprehensive treatment, we refer to \cite{martin_possibilistic_2025}. The case with non-vacuous prior information is covered in \cite{martin_valid_2023}, where the probability measure $P_\theta$ has to be replaced by a suitable outer measure.

\begin{algorithm}
\caption{The VIPER algorithm}
\label{alg:possibilistic IV}
\begin{algorithmic}[1]
\REQUIRE Data $W = \begin{bmatrix}Y & X\end{bmatrix}$ and $Z$; parameter value $\beta$; violation set $A$; approximation \texttt{type} $\in \{\chi^2, \text{MC}\}$
\STATE \COMMENT{Step 1: Maximum likelihood estimates}
\STATE $\hat{\Gamma} \gets (Z^\intercal Z)^{-1} Z^\intercal W$
\STATE $\Hat{\Psi} \gets n^{-1} (W - Z \Hat{\Gamma})^\intercal (W - Z \Hat{\Gamma})$
\STATE $\hat{\Sigma}(\beta) \gets R(\beta)^{-1}\, \hat{\Psi}\, [R(\beta)^{\intercal}]^{-1}$
\STATE \COMMENT{Step 2: Projection onto the violation set}
\STATE $t(\beta) \gets \Hat{\Gamma} \begin{bmatrix} 1 \\ -\beta \end{bmatrix}$
\STATE $\hat{\alpha}(\beta) \gets \mathrm{Proj}_A^{Z^{\intercal}Z}(t(\beta))$
\STATE \COMMENT{Step 3: Posterior possibility function}
\STATE $C_\beta \gets \sup_{\beta' \in \mathbb{R}} f_{\mathrm{S}}(\hat{\alpha}(\beta'), \beta', \hat{\Sigma}(\beta') \mid W)$
\STATE $f(\beta \mid \alpha \in A, W) \gets f_{\mathrm{S}}(\hat{\alpha}(\beta), \beta, \hat{\Sigma}(\beta) \mid W) / C_\beta$
\STATE \COMMENT{Step 4: Validification}
\IF{\texttt{type} == $\chi^2$}
    \STATE $\pi_w(\beta \mid A) \gets 1 - F(-2 \log f(\beta \mid \alpha \in A, W = w))$
\ELSIF{\texttt{type} == MC}
    \STATE $\Hat{\theta}(\beta, A) \gets (\hat{\alpha}(\beta), \beta, \hat{\Sigma}(\beta), \hat{\gamma}_2)$
    \STATE $W_j \sim P_{\hat{\theta}(\beta, A)}, \quad j = 1, \ldots, M$
    \STATE $\pi_w(\beta \mid A) \gets M^{-1} \sum_{j=1}^M 1\{ f(\beta \mid \alpha \in A, W =W_j) \leq f(\beta \mid \alpha \in A, W = w) \}$
\ENDIF
\RETURN $\pi_w(\beta \mid A)$
\end{algorithmic}
\end{algorithm}

Algorithm~\ref{alg:possibilistic IV} outlines the steps for a single posterior evaluation. For multiple posterior evaluations at different values of $\beta$, some of the computations can be shared across evaluations, including the maximum likelihood estimates (lines 2 and 3) and the normalising constant (line 9).

We close this section with two remarks discussing connections to alternative inferential perspectives.

\begin{remark}
    Under the vacuous prior $f(\Gamma, \Psi) = 1$, the conditional possibility $f(\beta \mid \alpha \in A, W)$ coincides with the constrained profile likelihood ratio for testing $H_0: \beta = \beta_0$ within $\Theta_A$, and the $\chi^2$ approximation recovers the standard Wilks-type confidence set whenever $\beta$ is point-identified within $\Theta_A$. The possibilistic formulation nonetheless offers several advantages over a purely frequentist treatment. First, the validified construction in Proposition~\ref{prop:validity} delivers finite-sample validity uniformly across the partially-identified regime, where Wilks' approximation no longer applies and specific corrections would otherwise be required. The ``flat top" of $\pi_w(\cdot \mid A)$ over the partial identification region is the correct representation of the available information, rather than an artefact. Second, because $\pi_w(\cdot \mid A)$ is an outer probability measure, it admits coherent lower and upper probability bounds for arbitrary hypotheses of interest, as illustrated in Tables~\ref{tab:lower_upper_AJR} and \ref{tab:lower_upper_schooling}, providing richer inferential summaries than a confidence set or $p$-value alone.
\end{remark}

\begin{remark}
    Under a rectangular violation set $A$, the partially identified region for $\beta$ is an interval $[\beta_L, \beta_U]$ whose endpoints are themselves identifiable functionals of the reduced-form parameters. One might therefore consider performing inference directly on the bounds $(\beta_L, \beta_U)$. While possible, this conflates two distinct sources of uncertainty: sampling uncertainty about where the bounds lie, and the epistemic non-identification of $\beta$ within the bounds. A posterior distribution on the bounds collapses both into a single, prior-dependent object, and converting the resulting credible statements into frequentist-valid statements about $\beta$ would require additional machinery. The possibilistic construction instead keeps the two separate: the ``flat top'' of $\pi_w(\cdot \mid A)$ over the identified region encodes genuine non-identification, while the decay of its tails reflects sampling uncertainty. Crucially, this is achieved while retaining the finite-sample coverage guarantee of Proposition~\ref{prop:validity}.
\end{remark}

 \section{Experiments}

\subsection{Simulation experiments}\label{sec:simexp}

First, we consider a toy example with a single instrument and varying instrument validity. The single instrument is generated from a standard Gaussian distribution. The residual pairs are simulated from a bivariate Gaussian with unit variances and correlation $\rho = 1/2$. Then, we generate pairs $(Y_i, X_i)$ from (\ref{eq:structural_equation}) with $\gamma_2 = 1, \beta = 1$, and varying invalidity parameter $\alpha \in \{0, 0.25,  0.5\}$. The outcome and treatment are centred so that we do not need to include an intercept.

Our performance criterion is the empirical coverage of a $95\%$ uncertainty interval for the treatment effect $\beta$. We consider VIPER under both the $\chi^2$ approximation and Monte Carlo (MC) sampling from the validified posterior possibility and different violation sets $A$, and compare its empirical coverage to those of naive two-stage least squares (TSLS), plausible generalised method of moments with a Gaussian prior (PGMM-g) \citep{chernozhukov_plausible_2025}, and BudgetIV \citep{penn_budgetiv_2025}. The results are given in Table \ref{tab:coverage} and the full details are provided in Appendix \ref{sec:details_experiments}. Code to reproduce our findings is available at \url{https://github.com/gregorsteiner/PossibilisticIV}.

\begin{table}[ht]
\centering
\caption{Empirical coverage of $95\%$ uncertainty intervals across $500$ simulated datasets of size $n =100$. The value closest to the nominal coverage in each column is printed in \textbf{bold}. The second best value is printed in \sBest{grey}, except when there is a tie for the best value. Median interval lengths are displayed in brackets, except for the MC variants.}
\label{tab:coverage}
\resizebox{\columnwidth}{!}{
\begin{tabular}{lccc}
\toprule
Method & \(\alpha = 0.0\) & \(\alpha = 0.25\) & \(\alpha = 0.5\) \\
\midrule
VIPER ($A = \{0\}, \chi^2$) & 0.928 [0.401] & 0.330 [0.365] & 0.006 [0.350] \\
VIPER ($A = \{0\}$, MC) & 0.930 [---] & 0.334 [---] & 0.006 [---] \\
VIPER ($A = [-0.5, 0.5], \chi^2$) & 1.000 [1.483] & \textbf{1.000 [1.478]} & 0.984 [1.459] \\
VIPER ($A = [-0.5, 0.5]$, MC) & 1.000 [---] & \textbf{1.000} [---] & \textbf{0.962} [---] \\
VIPER ($A = [0.0, 0.5], \chi^2$) & \textbf{0.956} [0.991] & \textbf{1.000} [0.946] & 0.984 [0.906] \\
VIPER ($A = [0.0, 0.5]$, MC) & 0.934 [---] & \textbf{1.000} [---] & \textbf{0.962} [---] \\
TSLS & \sBest{0.936} [0.393] & 0.280 [0.360] & 0.004 [0.343] \\
PGMM-g & 0.994 [0.558] & 0.574 [0.537] & 0.024 [0.528] \\
BudgetIV ($\tau = 0$) & 1.000 [0.694] & \textbf{1.000} [0.690] & 1.000 [0.684] \\
BudgetIV ($\tau = 0.5$) & 1.000 [1.715] & \textbf{1.000} [1.712] & 1.000 [1.689] \\
\bottomrule
\end{tabular}
}
\end{table}

For $A = \{0\}$, coverage falls as the true $\alpha$ moves away from zero. Widening $A$ to include the true $\alpha$ maintains coverage above the nominal level, but overly wide $A$ makes the intervals too conservative. The Monte Carlo-based possibility functions tend to be slightly better than the $\chi^2$ approximation in this setting. We believe the slight undercoverage for $A = \{0\}$ when $\alpha =0$ is due to approximation error. The only other method that maintains coverage above the nominal level is BudgetIV. However, it is overly conservative even when the true $\alpha = 0.5$.

In the second experiment, we consider $p=5$ instruments generated from a multivariate standard Gaussian, $s$ of which are invalid. The coefficient $\alpha$ is chosen such that the first $s$ components are $0.1$ and the remaining $p-s$ components are zero. We set $\gamma_2 = c \cdot (1, \ldots,1)$, where $c$ is chosen such that the treatment equation $R^2$ %**should we explain what this is? Especially non-econometricians may not know this terminology.. ``treatment equation $R^2$'' or ``treatment $R^2$''? Mentioned twice more in the Appendix** ** Good point **
is  $0.15$, thus the instrument strength is moderate. The treatment effect is again set to $\beta = 1$ and the residuals are generated as above. We consider a sample size of $n=200$. Now, we also include gIVBMA \citep{steiner_bayesian_2025}, LeakyIV \citep{watson_bounding_2024} and the confidence interval method (CIIV) \citep{windmeijer_confidence_2021}. The latter is one of the plurality rule based methods and is expected to perform well if $s \leq 2$.

\begin{table}[ht]
\centering
\caption{Empirical coverage of $95\%$ uncertainty intervals across $200$ simulated datasets of size $n =200$, where $s$ out of $p=5$ instruments are invalid with $\alpha_i = 0.1$. The value closest to the nominal coverage in each column is printed in \textbf{bold}, and the second best value is printed in \sBest{grey}, except when there is a tie for the best value. Median interval lengths are reported in brackets, except for the MC variants.}
\label{tab:coverage_multiple_instruments}
\resizebox{\columnwidth}{!}{
\begin{tabular}{lcccc}
\toprule
Method & \(s = 0\) & \(s = 2\) & \(s = 3\) & \(s = 5\) \\
\midrule
VIPER ($A = \{0\}, \chi^2$) & \sBest{0.955} [0.713] & 0.815 [0.686] & 0.580 [0.660] & 0.125 [0.597] \\
VIPER ($A = \{0\}$, MC) & 0.940 [---] & 0.750 [---] & 0.550 [---] & 0.345 [---] \\
VIPER ($A = [-0.1, 0.1]^p, \chi^2$) & 1.000 [1.463] & 0.965 [1.316] & 0.920 [1.267] & 0.780 [1.373] \\
VIPER ($A = [-0.1, 0.1]^p$, MC) & 0.995 [---] & \textbf{0.960} [---] & \textbf{0.960} [---] & \textbf{0.935} [---] \\
VIPER ($A = [0.0, 0.2]^p, \chi^2$) & 0.820 [1.579] & 0.965 [1.460] & \sBest{0.975} [1.332] & 0.995 [1.406] \\
VIPER ($A = [0.0, 0.2]^p$, MC) & 0.900 [---] & \textbf{0.960} [---] & \sBest{0.975} [---] & \sBest{0.990} [---] \\
TSLS & 0.935 [0.623] & 0.650 [0.590] & 0.385 [0.562] & 0.055 [0.541] \\
PGMM-g & 0.990 [0.887] & 0.895 [0.860] & 0.685 [0.831] & 0.235 [0.809] \\
gIVBMA & \textbf{0.950} [0.961] & 0.930 [0.973] & 0.835 [1.001] & 0.420 [0.725] \\
LeakyIV ($\tau = 0.2$) & 1.000 [49.479] & 1.000 [49.287] & 1.000 [49.918] & 1.000 [49.450] \\
BudgetIV ($b = 1, \tau = 0$) & 1.000 [4.347] & 1.000 [4.786] & 1.000 [4.821] & 1.000 [4.628] \\
BudgetIV ($b = 1, \tau = 0.2$) & 1.000 [7.877] & 1.000 [8.677] & 1.000 [8.770] & 1.000 [8.309] \\
CIIV & 0.915 [0.602] & 0.540 [0.563] & 0.360 [0.569] & 0.075 [0.547] \\
\bottomrule
\end{tabular}
}
\end{table}

Table \ref{tab:coverage_multiple_instruments} shows the results. For $s = 0$, all methods achieve good coverage. As more instruments become invalid, naive approaches, including VIPER with $A = \{0\}$, lose coverage. Our two variants with $A$ containing the true $\alpha$ maintain good coverage even when all instruments are invalid. However, when $\alpha$ lies on a corner of the hypercube (e.g., $A = [0.0, 0.2]^p$ with $s = 0$ or $A = [-0.1, 0.1]^p$ with $s = 5$), the intervals can be slightly too short, particularly for the $\chi^2$ approximation. LeakyIV and BudgetIV can also maintain coverage above the nominal level for $s=5$, but are overly conservative. In Appendix~\ref{sec:additional_experiments}, we present additional results for varying sample sizes and instrument strengths, where VIPER consistently provides good coverage.

VIPER widens the confidence sets appropriately when the instruments are not assumed to be valid a priori. This ensures valid inference even if no valid instruments exist. These sets may include a non-unique mode, reflecting partial identification. However, as shown in the experiments above, choosing $A$ too wide can make the confidence sets overly conservative and less useful in practice.  BudgetIV can similarly maintain good coverage, as it also relies on partial identification, yet it may not return a plausible set for certain hyperparameter specifications (see Appendix \ref{sec:details_experiments}). In contrast, our method always yields a posterior possibility function, even if it is sometimes very uninformative.

\subsection{The effect of institutions on economic growth}

We illustrate our method with an empirical application estimating the effect of institutions on economic output. The analysis uses the dataset of 64 countries originally compiled by \cite{acemoglu_colonial_2001} and reexamined by \cite{chernozhukov_plausible_2025} to demonstrate their quasi-Bayesian approach that allows for small violations of instrument exogeneity.

The outcome is log GDP per capita in 1995, with the main predictor being protection against expropriation, a proxy for institutional quality. A central challenge is endogeneity: institutions may affect income, but income may also influence  institutions. To address this, \cite{acemoglu_colonial_2001} use the mortality of early European settlers as an instrument, arguing that long-term settlement incentives shaped institutional quality. The instrument’s validity rests on the assumption that settler mortality is exogenous (conditional on covariates). Following \cite{chernozhukov_plausible_2025}, a direct effect of settler mortality should not be stronger than $0.1$ in absolute terms, if any, justifying $A = [-0.1, 0.1]$ as a reasonable violation set. We consider the specification with an intercept and (normalised) distance from the equator as exogenous control variables and (log) settler mortality as the sole instrument. We project out the covariates and run our analysis on the residuals resulting from that projection. 

\begin{figure}[ht]
    \centering
    \begin{subfigure}{\linewidth}
    \includegraphics[width=\linewidth]{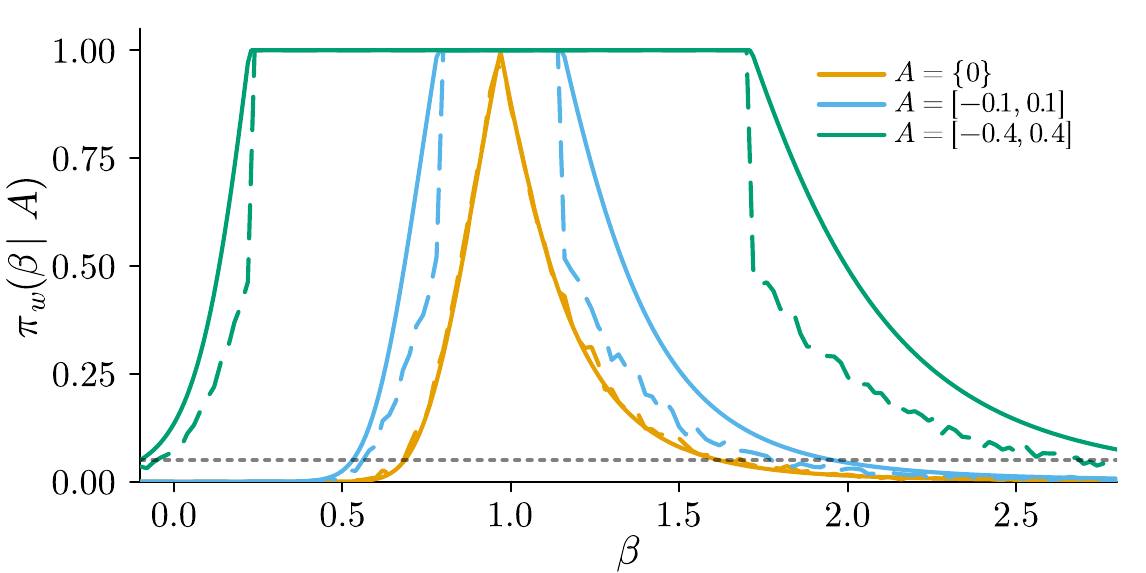}
    \caption{The effect of institutions on economic growth.}%, with $A = [-0.1, 0.1]$ and $A =[-0.4, 0.4]$.}
    \label{fig:AJR_results}
    \end{subfigure}
    \begin{subfigure}{\linewidth}
    \includegraphics[width=\linewidth]{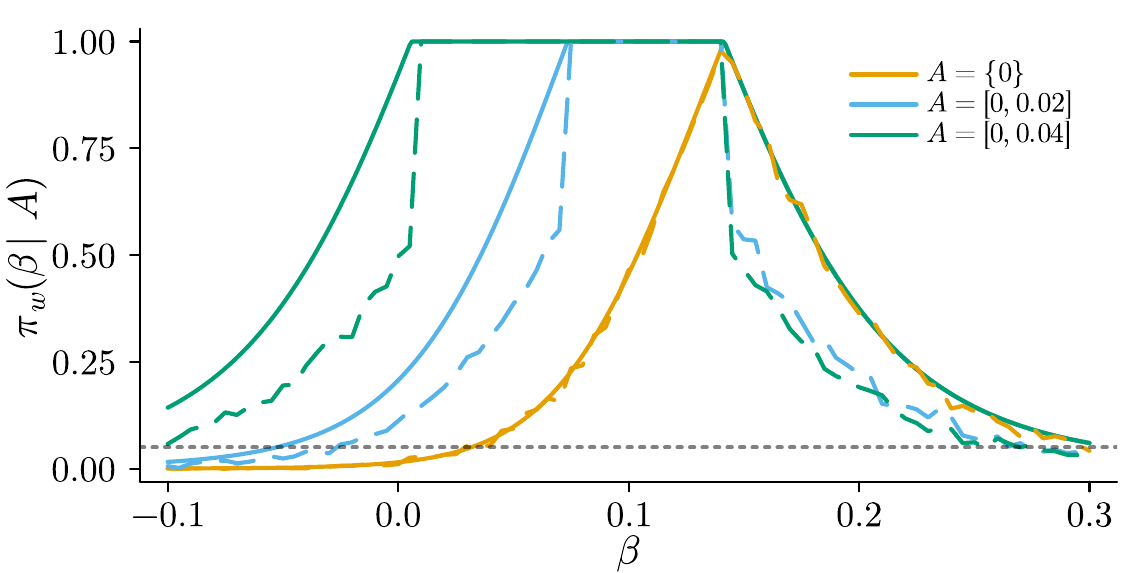}
    \caption{The returns to schooling.}%, with $A = [0, 0.02]$ and $A = [0, 0.04]$.}
    \label{fig:schooling_results}
    \end{subfigure}
    \caption{Validified posterior possibility functions under a perfectly valid instrument ($\alpha = 0$) and different potential violations. The solid line is based on the \(\chi^2\) approximation, while the dashed line displays the Monte Carlo approximation. The dashed grey line indicates the 0.05 level, such that the 95\% uncertainty interval for \(\beta\) includes all values where the posterior possibility function exceeds this threshold.}
\end{figure}

Figure \ref{fig:AJR_results} shows validified posterior possibility functions for the valid case and allowing some invalidity. The curve corresponding to $A = [-0.1, 0.1]$ is wider and lacks a unique mode, reflecting the partial identification of \(\beta\). Both indicate a significantly positive effect with heavier right tails, consistent with \cite{chernozhukov_plausible_2025}. For comparison, we also include the violation set $A = [-0.4, 0.4]$, which is wide enough for the significant effect to disappear. Table \ref{tab:intervals_AJR} shows that our uncertainty intervals are narrower than those of \cite{chernozhukov_plausible_2025}, especially on the right tail. Relaxing the exogeneity assumption and allowing for $\alpha \in [-0.1, 0.1]$ does not qualitatively change the conclusion that good institutions promote economic output.

For $A = \{0\}$, the $\chi^2$ and MC approximations look essentially identical. For larger $A$, however, the latter display steeper decay away from the partial identification region. This is expected as the Gaussian approximation cannot correctly capture the tail behaviour when $\beta$ is not point-identified. In these cases, we observe that the $\chi^2$ approximation leads to more conservative (but valid) inference. This suggests a practical strategy of fitting the $\chi^2$ approximation first and only computing the more expensive MC approximation if additional tightness is required.

\begin{table}[ht]
    \small
    \centering
    \caption{\textbf{The effect of institutions on economic growth:} $95\%$ uncertainty intervals for VIPER, TSLS, and PGMM \citep[taken from][]{chernozhukov_plausible_2025}. The intervals for our approach are based on the $\chi^2$ approximation. PGMM-u, PGMM-g and PGMM(d)-g refer to PGMM with, respectively, uniform prior, baseline Gaussian prior and diffuse Gaussian prior.}
    \label{tab:intervals_AJR}
    \begin{tabular}{lrr}
    \toprule
    Method & $95\%$ Interval\\
    \midrule
    VIPER ($A = \{0\}$) & $[0.69, 1.62]$\\
    VIPER ($A = [-0.1, 0.1]$) & $[0.53, 1.96]$ \\
    VIPER ($A = [-0.4, 0.4]$) & $[-0.10, 3.00]$ \\
    TSLS & $[0.56, 1.38]$ \\
    PGMM-u & $[0.58, 3.65]$ \\
    PGMM-g & $[0.49, 3.79]$\\
    PGMM(d)-g & $[0.22, 3.81]$\\
    \bottomrule
    \end{tabular}
\end{table}

\begin{table}
    \small
    \centering
    \caption{\textbf{The effect of institutions on economic growth:} Lower and upper probabilities for the hypothesis $\beta > 0$ under the constraint $\alpha \in A = [-a,a]$ (based on the MC approximation with $M = 10,000$).}
    \label{tab:lower_upper_AJR}
    %\begin{tabular}{lrr}
    %\toprule
    %A & Lower & Upper\\
    %\midrule
    %$\{0\}$ & 1 & 1\\
    %$[-0.1, 0.1]$ & 1 & 1\\
    %$[-0.2, 0.2]$ & 1 & 1\\
    %$[-0.3, 0.3]$ & 1 & 1\\
    %$[-0.4, 0.4]$ & 0.927 & 1\\
    %$[-0.5, 0.5]$ & 0.603 & 1\\
    %\bottomrule
    %\end{tabular}
    \begin{tabular}{lrrrrrr}
    \toprule
    $a$ & $0$ & $0.1$ & $0.2$ & $0.3$ & $0.4$ & $0.5$ \\
    \midrule
    Lower & 1     & 1       & 1       & 1       & 0.927   & 0.603   \\
    Upper & 1     & 1       & 1       & 1       & 1       & 1       \\
    \bottomrule
    \end{tabular}
\end{table}

To further interpret the obtained posterior possibility, we can view it as an upper bound on a precise (subjective) probability measure. This allows to deduce a corresponding lower bound, and therefore provides a probability interval for the event of interest. Table \ref{tab:lower_upper_AJR} displays such intervals for the hypothesis $\beta > 0$, given by the pair $\left( 1 - \sup_{\beta \leq 0} \pi_w(\beta \mid A), \sup_{\beta > 0} \pi_w(\beta \mid A) \right)$, under different choices of the violation set $A$. The lower probability drops below the nominal level only when $\alpha$ is close to 0.4 in absolute value. Thus, the qualitative effect of institutions on economic output is very robust to reasonable violations of the exogeneity assumptions.

\subsection{The returns to schooling}

We revisit the relationship between education and wages using the dataset from \cite{card_using_1995}. Education choices are not random, but are influenced by unobserved traits that also affect earnings, so the observed link between schooling and wages may not reflect the true causal effect. \cite{card_using_1995} addresses this endogeneity by using college proximity as an instrumental variable. The outcome is (the logarithm of) hourly wages and the treatment is years of schooling received, which we instrument with proximity to a four-year college. We partial out the available covariates, including experience, experience squared, family background, marital status, race, region, and parents’ education. The parental education variables are missing for a substantial proportion of the sample. Following \cite{card_using_1995}, we impute them using the mean and include an indicator for missingness.

We also expect the direct effect of proximity to a college on hourly wages to be positive, if it exists. A magnitude of the effect in between zero and $4\%$ seems reasonable to us. Thus, we consider the violation sets $A = [0, 0.02]$ and $A = [0, 0.04]$. Figure~\ref{fig:schooling_results} presents the results. For $A = \{0\}$, the posterior mode is just below $0.15$, comparable to the estimates in \cite{card_using_1995}, and the effect is significant, since the uncertainty intervals do not include zero. In the other two cases, the left tail is  shifted to the left and the effect is no longer significant, since the uncertainty intervals now include zero.  Thus, positive returns to education are rather  sensitive to violations of the exogeneity assumption of the instrument ``college proximity''. Finally, under the MC approximation, the right tail is thinner, assigning lower possibility to very high returns to schooling.

%\begin{figure}[h]
%    \centering
%    \includegraphics[width=\linewidth]{Schooling_Possibility_Contour.pdf}
%    \caption{\textbf{The returns to schooling:} Validified posterior possibility functions under a perfectly valid instrument ($\alpha = 0$) and allowing for potential violations $A = [0, 0.02]$ and $A = [0, 0.04]$. The solid line is based on the $\chi^2$ approximation, while the dashed line displays the Monte Carlo approximation. The dashed grey line indicates the 0.05 level, such that the 95\% uncertainty interval for $\beta$ includes all values where the posterior possibility function exceeds this threshold.}
%    \label{fig:schooling_results}
%\end{figure}

As in the previous example, we can compute lower and upper probabilities for the hypothesis that the returns to schooling are positive, see Table~\ref{tab:lower_upper_schooling}. We see that the positive treatment effect is robust to a direct effect of college proximity on wages of $1\%$, whereas larger violations lead the lower probability to drop significantly. For a direct effect of $4\%$, the lower probability of positive returns to schooling is only approximately $0.52$.

\begin{table}
    \small
    \centering
    \caption{\textbf{The returns to schooling:} Lower and upper probabilities for the hypothesis $\beta > 0$ under the constraint $\alpha \in A = [0,a]$ (based on the MC approximation with $M = 5,000$).}
    \label{tab:lower_upper_schooling}
    %\begin{tabular}{lrr}
    %\toprule
    %A & Lower & Upper\\
    %\midrule
    %$\{0\}$ & 1 & 1\\
    %$[0, 0.01]$ & 1 & 1\\
    %$[0, 0.02]$ & 0.887 & 1\\
    %$[0, 0.03]$ & 0.748 & 1\\
    %$[0, 0.04]$ & 0.517 & 1\\
    %\bottomrule
    %\end{tabular}
    \begin{tabular}{lrrrrr}
    \toprule
    $a$     & $0$ & $0.01$ & $0.02$ & $0.03$ & $0.04$ \\
    \midrule
    Lower & 1   & 1      & 0.887  & 0.748  & 0.517  \\
    Upper & 1   & 1      & 1      & 1      & 1      \\
    \bottomrule
    \end{tabular}
\end{table}

\section{Conclusion}

In this paper, we propose valid IV estimation with possibilistic endogeneity robustness (VIPER), a method that allows for valid inference under user-specified violations of the instrument exogeneity assumption. VIPER combines, within a single posterior-like uncertainty curve, the sampling uncertainty about the parameters and the epistemic uncertainty arising from partial identification induced by potential instrument invalidity. The inference reduces to a projection of a reduced-form statistic onto the violation set, followed by a validification step that calibrates the resulting uncertainty. Crucially, the exact validified construction enjoys a finite-sample coverage guarantee whenever the violation set contains the true violation parameter, a property that distinguishes it from existing methods relying on asymptotic arguments. It further supports sensitivity analysis as the violation set is varied, and performs well across simulations and two real-data applications.

\paragraph{Limitations and future work.} Several limitations point to directions for future work. First, inference is performed only relative to the user-specified set $A$, so its usefulness depends on domain knowledge about which instruments may be invalid and to what degree. If the specified set of violations is too large, the resulting inference will tend to be overly conservative, or even completely uninformative. A natural strategy is to report breakdown points where certain effects disappear, but with multiple instruments it is not obvious how to trade-off violations in different directions. Second, we consider a linear structural model with Gaussian errors and a single endogenous regressor. Extending these ideas to generalised linear models or nonparametric IV would be an interesting direction for future work. Finally, the more exact MC approximation is computationally demanding, which is why we report interval lengths only for the $\chi^2$ surrogate. Closing this gap, through intermediate procedures that improve on the $\chi^2$ surrogate while remaining substantially cheaper than full MC sampling, is a promising direction.

%\begin{contributions} % will be removed in pdf for initial submission 
					  % (without ‘accepted’ option in \documentclass)
                      % so you can already fill it to test with the
                      % ‘accepted’ class option
%    Briefly list author contributions. 
%    This is a nice way of making clear who did what and to give proper credit.
%    This section is optional.
%
%    H.~Q.~Bovik conceived the idea and wrote the paper.
%    Coauthor One created the code.
%    Coauthor Two created the figures.
%\end{contributions}

\begin{acknowledgements}
We thank Ryan Martin for helpful comments that substantially improved this paper. This research is supported by the Ministry of Education, Singapore, under its Academic Research Fund Tier 1 (RS02/24), and by the Singapore Ministry of Digital Development and Information under the AI Visiting Professorship Programme (AIVP-2024-004).
\end{acknowledgements}

\bibliography{references}

%Further instructions will be given when a paper is accepted.
%\vspace*{-10pt}

\newpage

\onecolumn

\title{Possibilistic Instrumental Variable Regression with Potentially Invalid Instruments \\(Supplementary Material)}
\maketitle

\appendix

\section{Derivations}

\renewcommand{\thefigure}{A.\arabic{figure}}
\renewcommand{\thetable}{A.\arabic{table}}
\renewcommand{\theequation}{A.\arabic{equation}}
\setcounter{figure}{0}
\setcounter{table}{0}
\setcounter{equation}{0}

\subsection{The structural posterior possibility function} \label{sec:structural_function}

This section derives the closed-form solution for the structural posterior possibility. We put completely uninformative prior possibility functions on $\Gamma$ and $\Psi$, that is $f(\Gamma, \Psi) = 1$. This reduces the problem to maximising the likelihood, or equivalently the log-likelihood $\log p(W \mid \Gamma, \Psi)$, under the constraint (\ref{eq:identification}) and the additional covariance equivalence. Given a value of $\Sigma$ and $\beta$, the reduced-form covariance $\Psi$ is fixed, so we just need to optimise with respect to $\Gamma$. To perform this optimisation, consider the log-likelihood
\begin{align*}
    \ell(\Gamma) = \log p(W \mid \Gamma, \Psi) = \mathrm{cst} - \frac{1}{2} \operatorname{tr} \left( \Psi^{-1} \left( \Gamma^\intercal (Z^\intercal Z) \Gamma - 2 \Gamma^\intercal Z^\intercal W \right) \right).
\end{align*}
Then, using that $Z^{\intercal}W = Z^{\intercal}Z\hat{\Gamma}$ and completing the square, we can rewrite $\log p(W \mid \Gamma, \Psi)$ as
\begin{align*}
     \log p(W \mid \Gamma, \Psi) = \mathrm{cst} - \frac{1}{2} \operatorname{tr} \left( \Psi^{-1} (\Gamma - \hat{\Gamma})^{\intercal} (Z^\intercal Z) (\Gamma - \hat{\Gamma}) \right)
\end{align*}
where $\Hat{\Gamma} = \begin{bmatrix}    \Hat{\gamma}_1 & \Hat{\gamma}_2 \end{bmatrix} = (Z^\intercal Z)^{-1} Z^\intercal W $ is the (unconstrained) least squares estimate. We maximise this log-likelihood in $\Gamma$ subject to the constraint
$$
\Gamma \begin{bmatrix}
    1 \\ -\beta
\end{bmatrix} - \alpha = 0.
$$
Solving this with Lagrange multipliers (LM) with a p-dimensional LM vector $\lambda$ we have the first-order conditions
\begin{align*}
    Z^\intercal Z \left(\Gamma - \Hat{\Gamma} \right) \Psi^{-1} &= \lambda \begin{bmatrix}
        1 \\ -\beta
    \end{bmatrix}^\intercal \\
    \Gamma \begin{bmatrix}
        1 \\ -\beta
    \end{bmatrix} - \alpha &= 0.
\end{align*}
Solving for $\Gamma$ yields
\begin{align*}
    \Gamma^*(\alpha, \beta, \Sigma) &= \Hat{\Gamma} + \left(\begin{bmatrix} 1 & -\beta    \end{bmatrix} \Psi \begin{bmatrix}
    1 \\ -\beta \end{bmatrix} \right)^{-1} \left(\alpha - \Hat{\Gamma} \begin{bmatrix} 1 \\ -\beta    \end{bmatrix} \right)    \begin{bmatrix} 1 & -\beta    \end{bmatrix}\Psi \\
    &= \Hat{\Gamma} + \frac{1}{\sigma_{11}} \left(\alpha - \Hat{\Gamma} \begin{bmatrix} 1 \\ -\beta    \end{bmatrix} \right)    \begin{bmatrix} 1 & 0    \end{bmatrix} \Sigma R(\beta)^\intercal
\end{align*}
where $\sigma_{11} = \begin{bmatrix} 1 & -\beta    \end{bmatrix} \Psi \begin{bmatrix}    1 \\ -\beta\end{bmatrix}$ is the marginal variance of $Y$ in the structural model (this follows from the relationship between the structural and reduced-form covariance). For completeness, the reduced form log-posterior possibility function is given by
\[
\log f_{\mathrm{RF}}(\Gamma, \Psi \mid W) = \log p(W \mid \Gamma, \Psi) - \log p(W \mid \hat{\Gamma}, \hat{\Psi}),
\]
where $\Hat{\Psi} = \frac{1}{n} (W - Z \Hat{\Gamma})^\intercal (W - Z \Hat{\Gamma})$ is the maximum-likelihood estimate of $\Psi$. Thus, the structural log-posterior possibility function is
\begin{align*}
    \log f_{\mathrm{S}}(\alpha, \beta, \Sigma \mid W) &= \log f_{\mathrm{RF}}(\Gamma^*(\alpha, \beta, \Sigma), R(\beta) \Sigma R(\beta)^\intercal \mid W) \\
    &= -\frac{n}{2} \log \lvert R(\beta) \Sigma R(\beta)^\intercal \rvert \\ &- \frac{1}{2} \text{tr} \left( (R(\beta) \Sigma R(\beta)^\intercal)^{-1} (W - Z \Gamma^*(\alpha, \beta, \Sigma))^\intercal (W - Z \Gamma^*(\alpha, \beta, \Sigma)) \right)
\end{align*}
To simplify this expression, define $M_Z = I_n - Z (Z^\intercal Z)^{-1} Z^\intercal$ and $t(\beta) = \hat{\Gamma} \begin{bmatrix} 1 \\ -\beta    \end{bmatrix}$, and note that $$W - Z\Gamma^*(\alpha, \beta, \Sigma) = M_ZW - \frac{1}{\sigma_{11}} Z (\alpha - t(\beta)) \begin{bmatrix} 1 & 0
\end{bmatrix} R(\beta)^{-1}.$$ Thus, we can express the structural log-posterior possibility function as
\begin{align}\label{eq:structural_possibility}
\begin{aligned}
    \log f_{\mathrm{S}}(\alpha, \beta, \Sigma \mid W)  &= -\frac{n}{2} \log \lvert R(\beta) \Sigma R(\beta)^\intercal \rvert \\ &- \frac{1}{2} \text{tr} \left( (R(\beta) \Sigma R(\beta)^\intercal)^{-1} W^\intercal M_Z W \right) \\ &- \frac{1}{2 } (\alpha - t(\beta))^\intercal \frac{Z^\intercal Z}{\sigma_{11}} (\alpha - t(\beta)).
\end{aligned}
\end{align}

\subsection{Proof of Proposition~\ref{prop:conditional_posterior} and computational considerations} \label{sec:conditional_beta}

Let $\bar{\mathbb{P}}_W$ be the outer measure corresponding to the joint posterior possibility $f_{\mathrm{S}}(\alpha,\beta, \Sigma \mid W)$. Then, the conditional outer measure of interest is
\begin{align*}
    \bar{\mathbb{Q}}_W(\beta \in B \mid \alpha \in A) = \frac{\bar{\mathbb{P}}_W(\alpha \in A, \beta \in B, \Sigma \in \mathbb{S}^2_{+})}{\bar{\mathbb{P}}_W(\alpha \in A, \beta \in \mathbb{R}, \Sigma \in \mathbb{S}^2_{+})}.
\end{align*}
The possibility function corresponding to $\bar{\mathbb{Q}}_W$ is
\begin{align} \label{eq:conditional_beta}
f(\beta \mid \alpha \in A, W)
    = \sup_{\alpha \in A,\, \Sigma \in \mathbb{S}^2_{+}}
      \frac{
          f_{\mathrm{S}}(\alpha, \beta, \Sigma \mid W)
      }{
          \sup_{\alpha' \in A,\, \beta' \in \mathbb{R},\, \Sigma' \in \mathbb{S}^2_{+}}
          f_{\mathrm{S}}(\alpha', \beta', \Sigma' \mid W)
      }.
\end{align}

To solve the optimisation problem in (\ref{eq:conditional_beta}), we marginalise out $\Sigma$ first to simplify the problem. To marginalise out $\Sigma$, we plug in  the maximum-likelihood estimator of the reduced-form covariance $\Psi$ given by
$$
\Hat{\Psi} = \frac{1}{n} (W - Z \Hat{\Gamma})^\intercal (W - Z \Hat{\Gamma})
$$
We can express $\hat{\Sigma}(\beta)$ as a function of $\beta$ in a way that
\begin{equation}
\label{eq:SigmaEstimate}
R(\beta) \hat{\Sigma}(\beta) R(\beta)^\intercal = \hat{\Psi};
\end{equation}
indeed, for any $\beta$, we can ensure that \eqref{eq:SigmaEstimate} holds by taking
\[
\hat{\Sigma}(\beta) = \begin{bmatrix}
    \hat{\Psi}_{11} - 2\beta \hat{\Psi}_{12} + \beta^2 \hat{\Psi}_{22} & \hat{\Psi}_{12} - \beta \hat{\Psi}_{22} \\
    \hat{\Psi}_{12} - \beta \hat{\Psi}_{22} & \hat{\Psi}_{22}
\end{bmatrix}.
\]
This ensures that whatever the considered value of $\beta$, we can still achieve the global maximum in $\Psi$. Fixing $\Sigma = \hat{\Sigma}(\beta)$, we obtain from equation (\ref{eq:structural_possibility})
\begin{align*}
    \log f_{\mathrm{S}}(\alpha,\beta \mid W) &= \log f_{\mathrm{S}}(\alpha, \beta, \hat{\Sigma}(\beta) \mid W) \\
    &= \mathrm{cst} - \frac{1}{2 } (\alpha - t(\beta))^\intercal \frac{Z^\intercal Z}{\Hat{\sigma}_{11}(\beta)} (\alpha - t(\beta)).
\end{align*}

The aim is to maximise $f_{\mathrm{S}}(\alpha,\beta \mid W)$, or equivalently $\log f_{\mathrm{S}}(\alpha,\beta \mid W)$, with respect to $\alpha$ given the constraint $\alpha \in A$. For a fixed $\beta$, it is sufficient to minimise $(\alpha - t(\beta))^{\intercal} (Z^\intercal Z) (\alpha - t(\beta))$ under the considered constraint $\alpha \in A$. Clearly, $t(\beta)$ is the minimiser whenever $t(\beta) \in A$. Otherwise, the minimiser is the point $\alpha^* \in A$ that is closest to $t(\beta)$ in the distance implied by $Z^\intercal Z$, i.e. the projection of $t(\beta)$ onto $A$ with respect to the metric induced by $Z^\intercal Z$. Equivalently, the solution is the point that minimises the Mahalanobis distance from a distribution with mean vector $t$ and covariance matrix $\Hat{\sigma}_{11}(\beta) (Z^\intercal Z)^{-1}$.

When $A$ is a rectangle, i.e. of the form $A = [a_1, b_1] \times \ldots \times [a_p, b_p]$, this is a standard quadratic programming problem, which we solve using the \texttt{JuMP.jl} package \citep{lubin_jump_2023}. If $Z^\intercal Z$ is diagonal, this simplifies to clipping component-wise.

Star-shaped violation sets \citep[as in][]{penn_budgetiv_2025} can be accommodated by expressing it as a finite union of rectangles $A = \cup_{j=1}^p R_j$. For a given value of $\beta$, we then have to find the closest point $\alpha^{(j)} \in R_j$ to $t(\beta)$ for all $j=1,\ldots, p$. The projection onto the star domain is then given by $$\alpha^*(\beta) = \arg\min_{\alpha \in \{\alpha^{(1)}, \ldots, \alpha^{(p)}\}} \quad (\alpha - t(\beta))^\intercal Z^\intercal Z (\alpha - t(\beta)).$$

Alternatively, we can also bound a norm of $\alpha$, that is, specify the constraint set as  $A_\tau = \left\{ \alpha : \lVert \alpha \rVert \leq \tau \right\}$, where the threshold $\tau$ is the maximum invalidity budget across all instruments \citep[similar to][]{watson_bounding_2024}. This may be easier to specify in some settings. For the $\ell_2$ norm, this turns the optimisation problem into a Ridge-type regression problem, and we have that the optimal $\alpha$ is
\begin{align*}
    \alpha = (Z^\intercal Z + \lambda I_p)^{-1} Z^\intercal Z \Hat{\Gamma} \begin{bmatrix} 1 \\ -\beta  \end{bmatrix},
\end{align*}
where $\lambda$ is the Lagrange multiplier corresponding to a specific threshold $\tau$. In practice, one has to find $\lambda$ such that $\lVert \alpha \rVert_2^2 =  \tau$. For the $\ell_1$ norm, the optimisation problem could be solved by a LARS-type \citep{efron_least_2004} algorithm, but we leave details for future work.

It remains to renormalise the resulting posterior by finding the maximal value of $\beta$ over $\mathbb{R}$. This is easily done numerically.

\subsection{Proof of Proposition \ref{prop:validity}} \label{sec:proof_validity}

Let $\alpha_0 \in \mathbb{R}^p$ denote the true data-generating value of $\alpha$. Then, the validified posterior possibility as a functional of $W$, $\pi_W(\beta \mid \{\alpha_0\})$, is a probability integral transform, and its distribution is stochastically greater or equal to a Uniform distribution on $[0, 1]$ \citep[see for example][Section 2.1]{casella_statistical_2002}. Thus, for any $\delta \in [0, 1]$, we have $P_{\theta} \left( \pi_W(\beta \mid \{\alpha_0 \}) \leq \delta \right) \leq \delta$. For any violation set that contains the true value, $A \supseteq \{\alpha_0\}$, the validified posterior becomes no more informative, i.e., $\pi_w(\beta \mid A) \geq \pi_w(\beta \mid \{\alpha_0\})$, such that 
\[
P_{\theta} \left( \pi_W(\beta \mid A) \leq \delta \right) \leq P_{\theta} \left( \pi_W(\beta \mid \{\alpha_0 \}) \leq \delta \right) \leq \delta.
\]
This inequality holds for all parameters $\theta$, and therefore also holds for the supremum over the set $\Theta_A$, giving the desired result.

\section{Additional details on the simulation experiments}

\renewcommand{\thefigure}{B.\arabic{figure}}
\renewcommand{\thetable}{B.\arabic{table}}
\setcounter{figure}{0}
\setcounter{table}{0}

\subsection{Implementation details}  \label{sec:details_experiments}

Here, we provide additional details on the implementation of our method and competing baselines in our simulation experiments. We evaluate the validified posterior possibility $\pi_w(\beta \mid A)$ at the true value of $\beta$, where our uncertainty intervals cover if $\pi_w(\beta \mid A) > 0.05$ at the data-generating value of $\beta$. To explicitly find the intervals, we numerically solve $\pi_w(\beta \mid A) = 0.05$. Subtracting the two solutions yields the interval length. However, numerically solving for the intervals becomes computationally infeasible for the MC approximation (even for relatively small $M$). Thus, we only report median interval lengths for the $\chi^2$-based VIPER variants.

For the plausible GMM (PGMM) estimator, we put a baseline Gaussian prior with identity prior covariance on the moment restriction. This puts all values of $\alpha$ considered (transformed to the moment restriction) well within the centre of that distribution. The CIIV results are based on the implementation available at \url{https://github.com/xlbristol/CIIV} with default settings. The gIVBMA method is implemented by \url{https://github.com/gregorsteiner/gIVBMA.jl} and we choose the hyper-$g/n$ prior specification.

To implement BudgetIV, we use the \texttt{budgetIVr} R package. Their method has two hyperparameters, $b$ and $\tau$, which specify that at least $b$  components of $\alpha$ cannot exceed the threshold $\tau$. We set $b = 1$ and select the budget $\tau$ analogously to the thresholds we use for our possibilistic approach. This means that at least one instrument must satisfy  $\lvert \alpha_i \rvert \leq \tau$. In the case of multiple instruments, setting $b=p=5$ would provide a fairer comparison with our approach, since then all instruments are restricted to be potentially invalid only up to degree $\tau$. However, this configuration rarely yields a feasible set for the treatment effect, i.e., one where the affine subspace intersects the partial identification region, in which case BudgetIV returns no set. This is likely caused by BudgetIV's reliance on the no measurement error (NOME) assumption —the requirement that sampling error in the treatment equation be negligible—which does not hold in our setting\footnote{We are grateful to an anonymous reviewer for pointing this out.}. In contrast, our approach still produces a posterior possibility function based on projection, where the mode corresponds to the value of $\beta$ that implies an $\alpha$ closest to the violation set $A$. For this reason, we maintain $b=1$ for BudgetIV, which then always returns a plausible set. Unsurprisingly, this additional flexibility renders BudgetIV very conservative in the multiple instrument case.

The LeakyIV method is implemented in the \texttt{leakyIV} R package. We use the vector $\tau$-exclusion version as it more closely resembles the violation set specification in VIPER. We find the method to be very unstable in the single instrument case, where we often obtain singular covariance matrices in some of the bootstrap samples (despite setting \texttt{approx = TRUE}). Therefore, we only use LeakyIV as a benchmark in the scenario with multiple instruments. We set $\tau = 0.2$ analogously to the upper limits of the VIPER violation sets and use $100$ bootstrap samples to construct $95\%$ confidence intervals.

\subsection{Additional results} \label{sec:additional_experiments}

Table~\ref{tab:coverage_alt_errors} presents results for a modified version of the first simulation experiment with skewed and heavy-tailed error distributions. For the Student-$t$ errors, we generate the error components from a $t$-distribution with $\nu$ degrees of freedom, scaled to have approximately unit variance: $u_j \sim t_\nu / \sqrt{(\nu - 2)/\nu}$. To maintain the correlation structure, we set $\epsilon = u_1, \eta = \rho u_1 + \sqrt{1 - \rho^2} u_2$, where $u_1$ and $u_2$ are independently generated from the scaled $t$-distribution. We choose $\nu =3$ resulting in significantly heavier tails than the Gaussian residuals considered in the main text. For the skewed-normal errors, we generate each base component as $u_j = z_j + (\chi^2_1 - 1) / \sqrt{8}$, where $z_j \sim \mathcal{N}(0, 1)$, $\chi^2_1$ is a chi-squared random variable with one degree of freedom. We then standardize each component to have unit variance and create correlation in the same manner as in the Student-$t$ case. The coverage results are not much affected by the introduction of skewness, but improve when we consider data with heavier tails for $\alpha>0$, in the sense that the severe undercoverage found for  some methods in Table \ref{tab:coverage} is somewhat mitigated. This is because all methods tend to yield wider intervals under the Student-$t$ errors.

\begin{table}[ht]
\centering
\caption{Empirical coverage of $95\%$ uncertainty intervals across $200$ simulated datasets of size $n=100$ under alternative error distributions. The values closest to the nominal coverage are printed in bold. Median interval lengths are reported in brackets.}
\label{tab:coverage_alt_errors}
\begin{tabular}{lcccccc}
\toprule
Method & \multicolumn{3}{c}{SkewNormal} & \multicolumn{3}{c}{t(3)} \\
& \(\alpha=0.0\) & \(\alpha=0.25\) & \(\alpha=0.5\) & \(\alpha=0.0\) & \(\alpha=0.25\) & \(\alpha=0.5\) \\
\midrule
VIPER ($A = \{0\}, \chi^2$) & \textbf{0.955} [0.419] & 0.350 [0.368] & 0.000 [0.344] & 0.935 [1.282] & 0.865 [1.136] & 0.555 [1.062] \\
VIPER ($A = \{0\}$, MC) & 0.965 [---] & 0.355 [---] & 0.000 [---] & 0.945 [---] & 0.860 [---] & 0.580 [---] \\
VIPER ($A = [-0.5, 0.5], \chi^2$) & 1.000 [1.535] & \textbf{1.000} [1.487] & 0.975 [1.462] & 1.000 [2.383] & 1.000 [2.274] & \textbf{0.950} [2.070] \\
VIPER ($A = [-0.5, 0.5]$, MC) & 1.000 [---] & \textbf{1.000} [---] & 0.960 [---] & 1.000 [---] & \textbf{0.990} [---] & 0.935 [---] \\
VIPER ($A = [0.0, 0.5], \chi^2$) & 0.970 [1.026] & \textbf{1.000} [0.956] & 0.975 [0.905] & 0.975 [1.935] & 1.000 [1.807] & \textbf{0.950} [1.706] \\
VIPER ($A = [0.0, 0.5]$, MC) & \textbf{0.945} [---] & \textbf{1.000} [---] & \textbf{0.955} [---] & 0.945 [---] & \textbf{0.990} [---] & 0.935 [---] \\
TSLS & \textbf{0.955} [0.410] & 0.300 [0.361] & 0.000 [0.340] & \textbf{0.950} [1.119] & 0.775 [1.058] & 0.430 [0.965] \\
PGMM-g & 1.000 [0.581] & 0.565 [0.538] & 0.015 [0.512] & 0.960 [1.161] & 0.825 [1.042] & 0.480 [0.981] \\
BudgetIV ($\tau = 0$) & 1.000 [0.695] & \textbf{1.000} [0.680] & 1.000 [0.693] & 1.000 [1.942] & 1.000 [1.741] & 1.000 [1.963] \\
BudgetIV ($\tau = 0.5$) & 1.000 [1.716] & \textbf{1.000} [1.684] & 1.000 [1.708] & 1.000 [2.985] & 1.000 [2.743] & 1.000 [2.983] \\
\bottomrule
\end{tabular}
\end{table}

\begin{figure}
    \centering
    \includegraphics[width=0.9\linewidth]{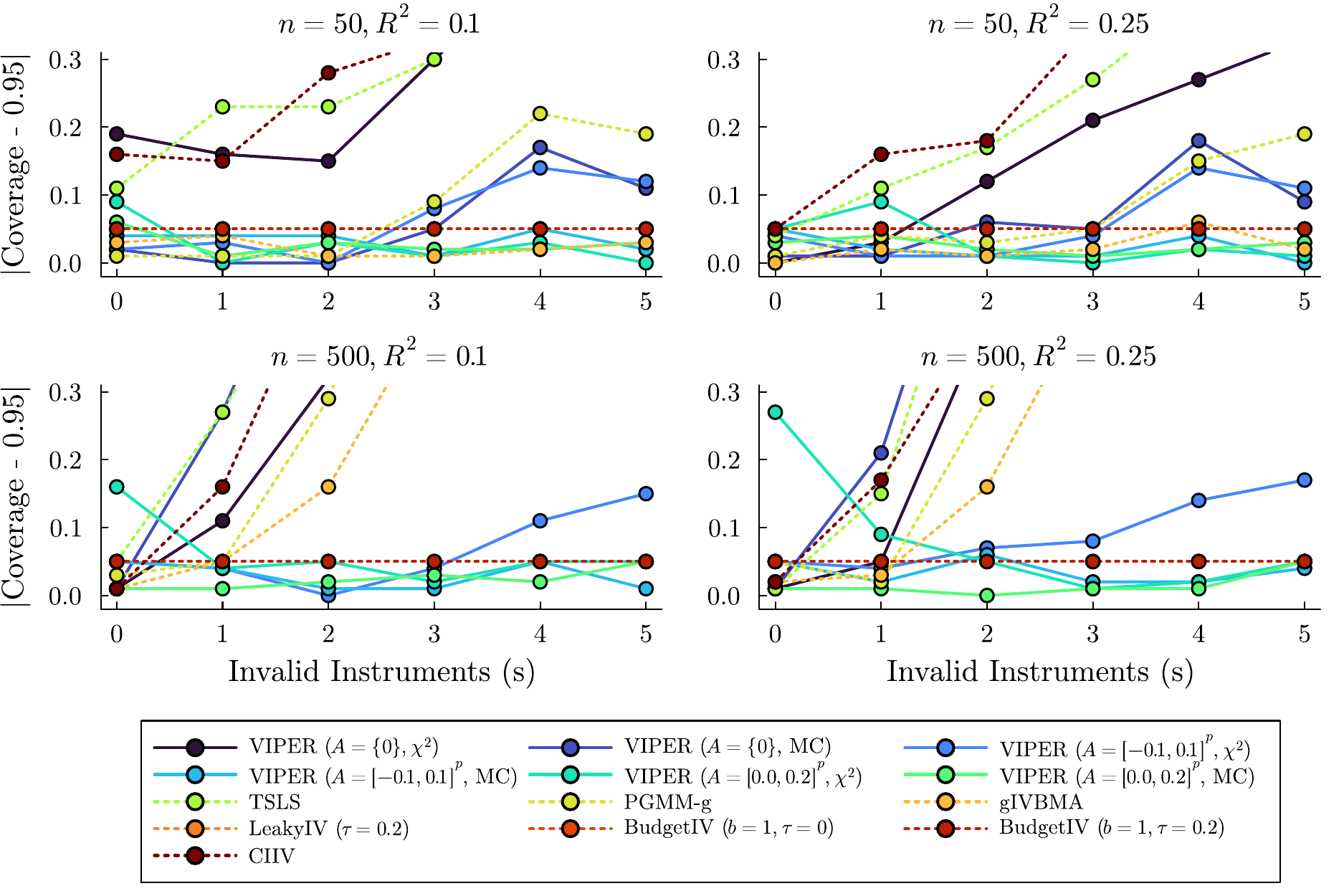}
    \caption{Absolute coverage error (lower is better) across $200$ simulated datasets for varying sample size ($n \in \{50, 500\}$), instrument strength (treatment equation $R^2 \in \{0.1, 0.25\})$, and number of invalid instruments $s$. The largest errors (above $0.3$) are truncated for better readability. Note that the lines for LeakyIV and both BudgetIV variants overlap.}
    \label{fig:sim_add_results}
\end{figure}

Figure~\ref{fig:sim_add_results} presents additional results for the multiple instrument experiment (the second experiment in Subsection \ref{sec:simexp}) with varying sample size ($n$) and instrument strength (as measured by the treatment equation $R^2$). Here, the y-axis displays the absolute deviation from the nominal coverage level (95\%) rather than the raw empirical coverage. The VIPER variants with an appropriate violation set, using the MC approximation, achieve the lowest coverage error in almost all scenarios. Only in the small sample with strong instruments is gIVBMA slightly superior. Of the non-VIPER methods, CIIV and TSLS display the worst overall performance. 
Despite being designed for this setting, PGMM-g consistently performs worse than gIVBMA. LeakyIV and BudgetIV never make large coverage errors (as they tend to be overly conservative, in which case the error is bounded by 0.05), but are still dominated by VIPER with $A=[0.0,0.2]^p$ and the MC approximation.

%\newpage

\end{document}